\documentclass[fleqn,10pt]{wlscirep}
\usepackage[utf8]{inputenc}
\usepackage[T1]{fontenc}
\usepackage{siunitx}
\title{
Optical Mode Sorting with a Programmable Diffractive Neural Network 
}   

\author[1,*,+]{Qian Zhang}
\author[1,+]{Shiyue Chen}
\author[1]{Juergen W Czarske}

\affil[1]{Laboratory of Measurement and Sensor Systems, Reinhart Koselleck Group, TU Dresden, Helmholtzstrasse 18, 01069, Dresden, Sachsen, Germany}
\affil[*]{qian.zhang@email.example}

\affil[+]{these authors contributed equally to this work}

\begin{abstract}

Programmable diffractive optical processors are particularly attractive for spatial light manipulation because their optical transformations can be dynamically reconfigured and adapted without modifying the physical hardware.
However, their practical performance is often limited by the gap between simulation and experiment caused by optical aberrations, alignment errors, and nonideal phase responses.
In this paper, we introduce a hybrid optimization framework that combines high-dimensional numerical design with low-dimensional hardware-in-the-loop calibration, enabling programmable diffractive optical networks to compensate experimentally for mismatch without retraining their underlying optical transformations.
We demonstrate a programmable optical diffractive neural network (ODNN) designed by back-propagation to spatially sort six linearly polarized modes supported by a multimode fiber.
The experimental distortions are represented using a truncated Zernike basis with only 19 correction coefficients per layer. These coefficients are optimized directly on the physical system using stochastic parallel gradient descent, avoiding re-optimization of the full pixelated phase masks.
Experimentally, the proposed calibration yields an SNR improvement of approximately 3.26 dB.
This separation of high-dimensional optical-function design from low-dimensional physical calibration provides a scalable route towards adaptive and reconfigurable spatial-mode processors for optical router on spatial modes and wavelengths, programmable photonic computer systems and quantum information processing.

\end{abstract}
\begin{document}

\flushbottom
\maketitle
%
%
\thispagestyle{empty}

\noindent 

\section*{Introduction}

Spatial-mode sorting is a key functionality in mode-division multiplexing, optical communications, high-dimensional information processing~\cite{ryf2012space}, and quantum key distribution~\cite{zhang2026space}.
Multi-plane light conversion (MPLC) provides a powerful approach for realizing such spatial transformations by cascading multiple phase modulation planes and free-space propagation~\cite{fontaine2019laguerre,kupianskyi2023high,pohle2023intelligent, rothe2025unlocking}.
When implemented using spatial light modulators (SLMs), these systems offer high reconfigurability and enable numerically optimized phase profiles to be directly deployed onto physical optical hardware~\cite{yang2023review}.
Beyond conventional high-channel-count mode multiplexing, multi-plane programmable phase modulation has also been exploited for reconfigurable optical switching and routing, enabling programmable unicast and multicast connectivity as well as joint space–wavelength routing \cite{Dinc2024}.
Such SLM-based routers are particularly attractive because optical connectivity can be reconfigured electronically by updating the phase patterns, eliminating mechanically moving components while enabling flexible allocation~\cite{cheng2018photonic}.
Moreover, SLMs provide a natural interface for closed-loop adaptive control: experimentally measured outputs can be used as feedback to update the phase profiles and compensate for modal coupling, misalignment, aberrations, and environmentally induced changes in few- and multimode fibers (MMFs) \cite{bao2025real,a2025self}.
Such adaptability is particularly important in quantum photonic systems, where unknown quantum states cannot be arbitrarily amplified, copied, or regenerated. Closed-loop wavefront correction using adaptive SLMs therefore provides a promising means to compensate distortions prior to detection, with particular relevance to high-dimensional quantum communication and quantum key distribution (QKD)~\cite{lib2025high}.

However, phase distributions optimized with ideal numerical models often suffer from a noticeable simulation-to-experiment gap once implemented experimentally. This discrepancy arises from various nonidealities in the physical system, including optical aberrations, imperfect alignment, off-axis arrangement, inaccurate propagation distances, non-planar wavefront, nonuniform SLM phase responses, and other systematic errors that are difficult to model precisely~\cite{kupianskyi2023high,a2025self}. A straightforward solution is to directly optimize the individual SLM pixels using experimentally measured outputs as feedback, but such pixel-wise hardware-in-the-loop optimization involves an extremely high-dimensional parameter space, which depends on the number of trainable parameters, and therefore requires a large number of measurements and iterations. An alternative approach is to minimize the simulation-to-experiment discrepancy by precisely aligning the positions of the phase masks~\cite{lib2025building}. However, alignment alone cannot compensate for wavefront distortions introduced by the folded optical architecture, such as those arising from repeated reflections, off-axis propagation, and surface non-planarity.

In this paper, we introduce a low-dimensional hardware-in-the-loop optimization strategy in which the experimentally induced wavefront distortion is parameterized using a limited set of Zernike polynomials\cite{noll1976zernike,xue2018adaptive}. Rather than re-optimizing the entire diffractive phase profile, only the corresponding Zernike coefficients are iteratively adjusted according to the experimentally measured mode-sorting performance. This physics-informed parameterization substantially reduces the dimensionality and computational complexity of the optimization while efficiently compensating for dominant low-order aberrations and systematic experimental errors, thereby bridging the gap between numerically designed and experimentally realized diffractive optical processors~\cite{Pohle2023,bearne2026diffractive,bao2025real}.
From a neural-network perspective, the phase profiles of an MPLC can also be regarded as trainable optical parameters, leading to the concept of optical diffractive neural networks (ODNNs), in which the diffractive layers are optimized using data-driven or gradient-based learning algorithms for a specific optical information-processing task~\cite{bearne2026diffractive, jia2023vector, zhang2022polarized,wang2025deep,rahman2024integration}.
MMF-based communciation systems represent a particularly relevant application for programmable diffractive mode processors, since their practical performance is strongly affected by modal coupling, environmental perturbations, and imperfections in the coupling between free-space optics and the fiber~\cite{xiong2018complete,czarske2016transmission}. In particular, lateral and angular misalignment, defocus, and wavefront aberrations during incoupling and outcoupling can alter the excited modal composition and degrade the performance of a numerically designed mode sorter. A programmable SLM-based ODNN provides a natural platform for compensating these deviations, because its phase response can be adaptively updated using experimentally measured feedback without modifying the optical hardware.
In this work, we therefore introduced a spatial mode sorter for a MMF with the programmable diffractive processor.
We design the mode sorter digitally and experimentally optimize the mode sorter using proposed low-dimensional algorithm. 
After hardware-in-the-loop optimization, the experimentally measured SNR is improved by approximately $3.26$~dB, indicating a substantial reduction of modal leakage and improved discrimination between the desired output channels.
The total insertion loss of the complete system is 8.14 dB. 
The amplitude error of each mode decreased from 0.14 to 0.10. These results demonstrate an efficient and experimentally practical route for calibrating reconfigurable diffractive optical processors without requiring high-dimensional pixel-wise optimization.


\section*{Experimental setup for realizing ODNN}

\begin{figure}[h]
    \centering
    \includegraphics[width=0.8\linewidth]{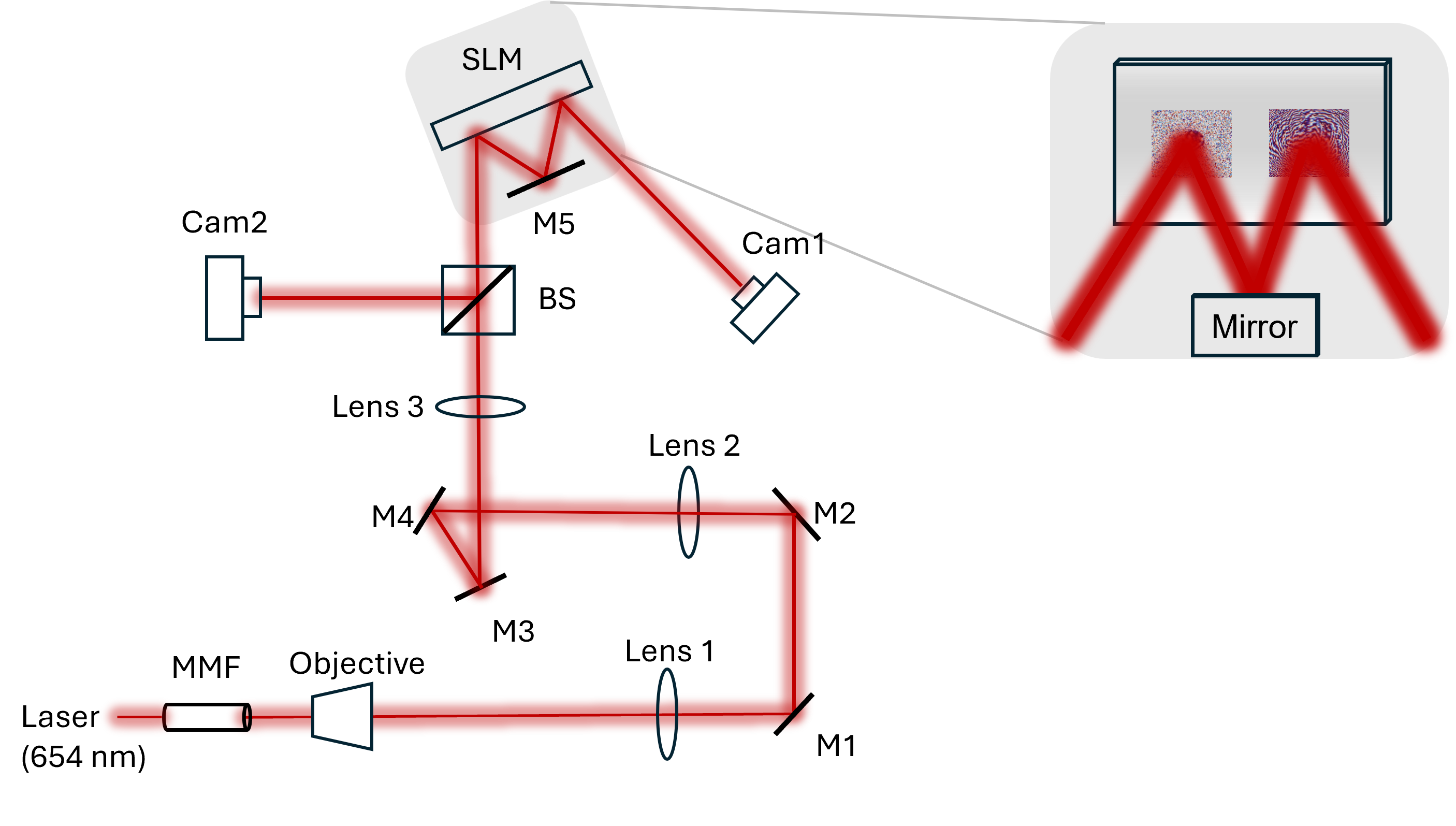}
    \caption{Schematic of the ODNN experimental setup.
    An SLM-based multi-plane optical architecture is employed as the physical platform for implementing the ODNN. BS denotes the beam splitter, and CAM1 and CAM2 denote the two camera branches used for system observation.}
    \label{fig:experimental-optical-setup}
\end{figure}

Figure~\ref{fig:experimental-optical-setup} illustrates the experimental setup of the reconfigurable MPLC system. 
For the mode-sorting experiment, a graded-index MMF with a numerical aperture of 0.1 and a core diameter of \SI{10}{\micro\metre} was employed. At the operating wavelength of \SI{654}{\nano\metre}, the fiber supports six spatial modes for each polarization, namely $\mathrm{LP}_{01}$, $\mathrm{LP}_{11a}$, $\mathrm{LP}_{11b}$, $\mathrm{LP}_{21a}$, $\mathrm{LP}_{21b}$, and $\mathrm{LP}_{02}$. The output field from the MMF was subsequently directed to the SLM-based ODNN for spatial-mode sorting. Since the phase modulation of the SLM is polarization dependent, a polarizer and a half-wave plate were inserted before the SLM. The polarizer selects a well-defined linear polarization, while the half-wave plate rotates its orientation to match the polarization axis required for efficient phase modulation by the SLM. This ensures that the incident field experiences the calibrated phase response of the SLM throughout the multi-plane transformation.
For the application in quantum scenarios, where polarization can not be touched before the measurement, phase-type MEMS SLM can be applied~\cite{rocha2024fast}. 
The multi-plane transformation is implemented using a single phase-only SLM (Hamamatsu photonics, X13138-07), on which two independently optimized phase profiles are displayed at spatially separated regions. By folding the optical path between the SLM and a mirror, the beam sequentially interacts with several phase-modulation regions, thereby realizing a multiple-plane optical transformation using a single programmable device. 
After the final phase modulation, the output intensity distribution is recorded by a camera (CAM 1), which is used to evaluate the mode-sorting performance and provides the experimental feedback for subsequent hardware-in-the-loop optimization. A beam splitter (BS) directs part of the optical signal to CAM2 for auxiliary monitoring of the optical field and system stability. This SLM-based folded architecture enables the phase profiles of all modulation planes to be dynamically updated without modifying the physical optical setup, providing a fully reconfigurable platform for both numerical phase-mask implementation and experimental optimization. 

Based on the physical parameters of the SLM and the propagation geometry of the two-plane folded optical system, we numerically designed a two-layer diffractive mode sorter for the six supported LP modes. The experimentally relevant parameters, including the SLM sampling and the propagation distances between successive phase-modulation planes, were incorporated into the numerical propagation model.  
The ODNN was modeled using the Angular Spectrum Method (ASM), and the phase distributions of the diffractive layers were optimized via back-propagation within a neural network framework.
The phase profiles of the two diffractive layers were then jointly optimized to transform each input mode into a predefined spatial output region with minimum power leakage to the other channels.
After numerical optimization, the resulting phase masks were directly loaded onto the corresponding regions of the SLM and experimentally evaluated using the setup shown in Fig.~\ref{fig:experimental-optical-setup}.

\section*{Design of the mode sorter}

Based on the physical parameters of the SLM and the propagation geometry of the folded optical system, we constructed a numerical model of the two-plane programmable diffractive mode sorter. The model was implemented at the operating wavelength of $\lambda=654$~nm and reproduced the pixel size of the SLM as well as the propagation distances between successive phase-modulation planes. 
The angular spectrum method (ASM) is used to simulate the free-space propagation between adjacent planes, which is described by a numerical propagation operator $\mathcal{P}_{z}$, such that the optical field after the $l-$th layer can be written as:
\begin{equation}
E_{l+1}(x,y)
=
\mathcal{P}_{z_l}
\left\{
E_l(x,y)
\exp\left[i\phi_l(x,y)\right]
\right\},
\end{equation}
where $\phi_l(x,y)$ denotes the phase profile of the $l$th diffractive layer and $z_l$ is the corresponding propagation distance.

The six spatial modes supported by the MMF were used as the input basis of the mode sorter. Each input mode was assigned to an individual detection region at the output plane. The phase profiles were jointly optimized such that the optical power of each input mode was concentrated within its designated output region while leakage into the remaining channels was suppressed. 
For a given input mode $i$, the optical power detected within output region $j$ is denoted by $P_{ij}$. The numerical optimization was performed by maximizing the power in the desired channel while minimizing the power coupled to undesired output regions. In this way, all six input modes were optimized simultaneously rather than designing an independent phase profile for each mode.

The ODNN was designed according to the physical geometry of the experimental system. It comprises two phase-only diffractive layers, each containing $285\times285$ independently trainable pixels, corresponding to a total of $243{,}675$ trainable phase elements (neurons) across the network. The separation between adjacent diffractive layers is $4.1~\mathrm{cm}$. 
The diffractive phase profiles were jointly optimized using the Adam optimizer~\cite{kingma2015adam}, with the mean squared error (MSE) between the predicted and target intensity distributions as the loss function. During training, the phase values of both diffractive layers were treated as trainable parameters and updated by back-propagation through the optical propagation model. After convergence, the optimized phase distributions were wrapped to the $0$--$2\pi$ range and converted into grayscale patterns according to the calibrated phase response of the SLM.
To quantitatively characterize the performance of the designed ODNN, the signal-to-noise ratio (SNR), amplitude error, and insertion loss were employed as evaluation metrics. The influence of finite phase resolution was then investigated numerically by varying the phase discretization level of the SLM. The resulting performance trends are presented in the Results section.

\begin{figure}[htbp]
    \centering
    \includegraphics[width=0.99\textwidth]{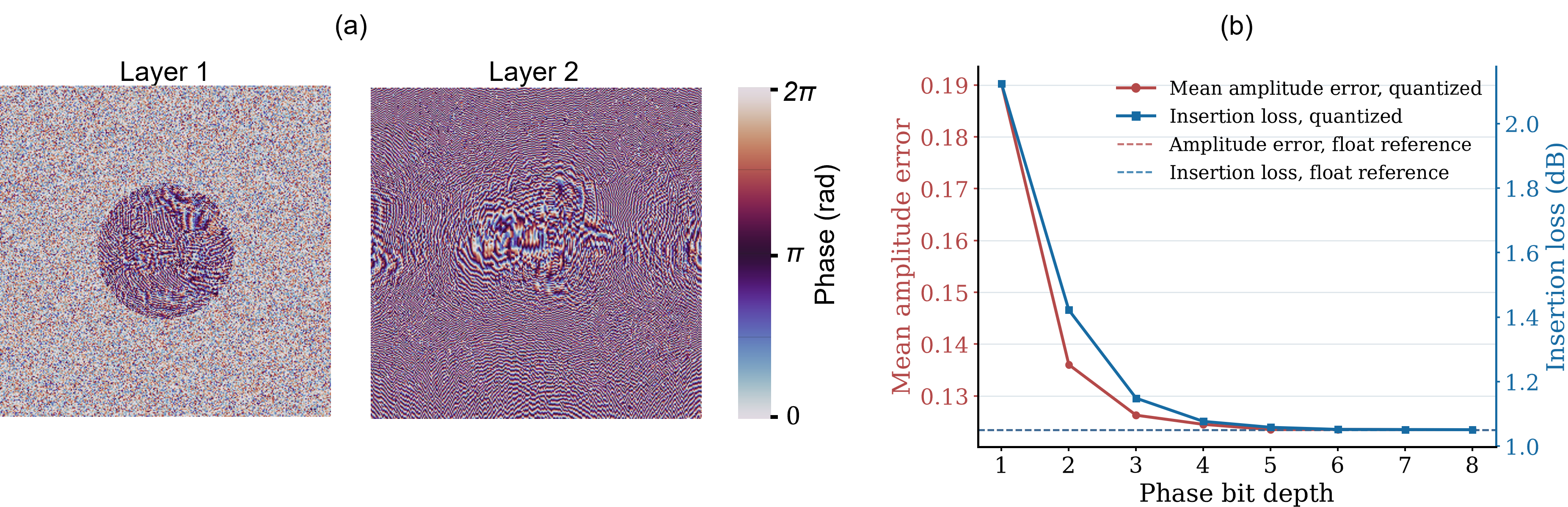}
    \caption{Numerical design and phase-quantization analysis. (a) Optimized phase masks of the two diffractive layers. (b) Simulated SNR and mean amplitude error as functions of phase bit depth. The dashed lines indicate the continuous-phase results. In the case of 5-bits discretization, the insertion loss in the simulation is about 1.1 dB.}
    \label{fig:results_sim}
\end{figure}

\section*{Optimization with Zernike coefficients}

The numerical design provides the high-dimensional phase structure required for the target mode transformation. However, the phase profiles are optimized using a nominal propagation model and therefore do not account for experimental imperfections, such as optical aberrations, alignment errors, propagation-distance uncertainties, and nonideal SLM responses~\cite{philipp2019diffraction,gao2021distortion,radner2021fieldprogrammable,bilsing2022imaging}. 
Direct pixel-wise optimization of the SLM can compensate for these discrepancies, but requires optimizing a very large number of degrees of freedom. To reduce the dimensionality of the experimental calibration, the residual phase correction applied to each diffractive layer is parameterized using a truncated set of Zernike modes rather than optimized pixel by pixel~\cite{schmieder2022twowavelength}.
For the $l$th diffractive plane, the correction phase is expressed as
\begin{equation}
\phi_{\mathrm{Z}}^{(l)}(\rho,\theta)
= \sum_{k=1}^{19} a_{k}^{(l)} Z_k(\rho,\theta),
\end{equation}
where $Z_k(\rho,\theta)$ denotes the $k$th Zernike polynomial defined over the normalized circular aperture, and $a_k^{(l)}$ is its corresponding coefficient. The coordinates $\rho$ and $\theta$ denote the normalized radial and azimuthal coordinates, respectively. In this work, 19 Zernike modes from $Z_1$ to $Z_{19}$ are considered. 
The piston term $Z_0$ is fixed to zero because a spatially uniform phase offset does not affect the measured intensity distribution, leaving 19 effective optimization parameters for each diffractive layer. The selected Zernike basis includes both low- and higher-order aberration terms, including tip, tilt, defocus, astigmatism, coma, and spherical aberration.
The phase pattern displayed on the SLM is obtained by superimposing the experimentally optimized Zernike correction onto the numerically designed diffractive phase,
\begin{equation}
\phi_{\mathrm{SLM}}^{(l)}
= \phi_{\mathrm{DNN}}^{(l)} + \phi_{\mathrm{Z}}^{(l)}.
\end{equation}
where $\phi_{\mathrm{DNN}}^{(l)}$ represents the original phase profile obtained from numerical optimization. In this way, the high-dimensional phase structure responsible for the desired mode transformation is retained, while the experimentally induced residual wavefront errors are compensated within a substantially reduced parameter space. 

The Zernike coefficients were optimized directly on the experimental setup using stochastic parallel gradient descent (SPGD)~\cite{vorontsov1998stochastic}. 
At each iteration, each Zernike coefficient was randomly varied by a positive or negative value, and the corresponding output intensity distributions were recorded by CAM1. The experimentally measured mode-sorting performance was used as the feedback signal to determine the coefficient update. 
In the experiment, the corresponding output intensity distributions were recorded by CAM1, and the contrast between the designated monitoring regions and the background region (SNR) was used to evaluate the loss. The perturbation amplitude and learning rate were gradually decreased as the optimization progressed, allowing coarse exploration during the initial iterations and fine correction near convergence.
This optimization constitutes a direct hardware-in-the-loop procedure: the SLM applies the perturbed phase distribution, the physical optical system performs the forward propagation, and the experimentally measured mode-sorting performance provides the feedback for the subsequent update. At each iteration, two oppositely perturbed Zernike-coefficient states are experimentally evaluated, yielding two corresponding loss values that are used to determine the SPGD update direction, independent of the number of optimized Zernike coefficients~\cite{vorontsov1998stochastic}.
Consequently, an accurate differentiable model of the experimental system is not required, and systematic deviations that are difficult to represent in numerical simulations can be compensated directly in the physical system. In addition, the Zernike basis describes the dominant smooth wavefront distortions using only a small number of physically interpretable parameters, substantially reducing the search space compared with pixel-wise hardware optimization~\cite{zhu2020automated,xue2018adaptive,segel2016modal}.


\section*{Results}

\begin{figure}[t]
    \centering
    \includegraphics[width=\linewidth]{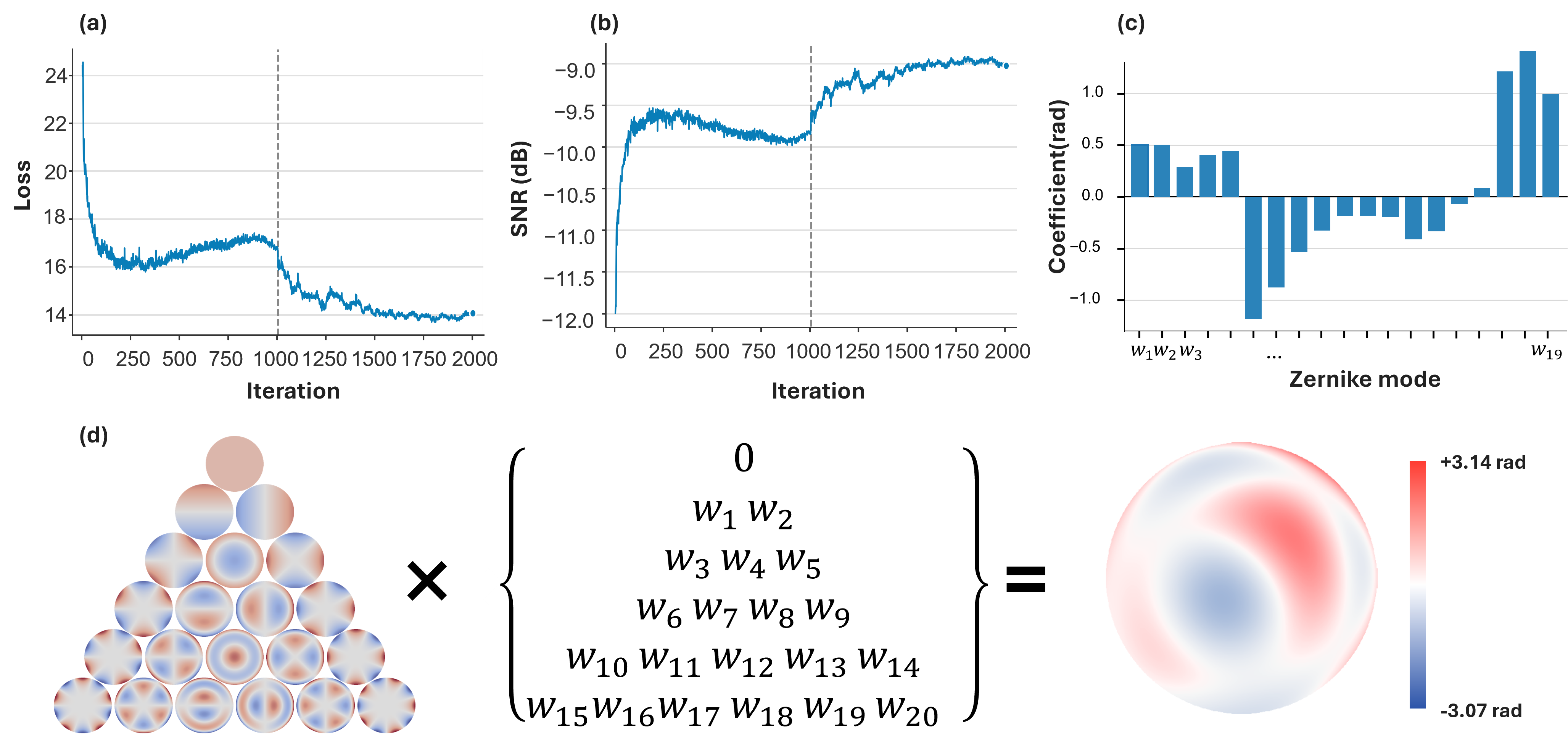}
    \caption{Representative single-layer optimization at $654$~nm. (a) Loss  during SPGD optimization. (b) SNR during SPGD optimization. (c) Optimized Zernike coefficients for the first layer. (d) Zernike modes and the reconstructed residual phase for the first layer.}
    \label{fig:optimized_residual_phase}
\end{figure}

First,the ODNN was tested in the simualtion and the influence of phase quantization on the mode-sorting performance was also evaluated numerically, as shown in Fig.~\ref{fig:results_sim}. 
At low bit depths, phase quantization results in a lower SNR and a larger mean amplitude error. As the phase resolution increases, both metrics rapidly approach their continuous-phase values. At a resolution of at least 5 bits, corresponding to 32 phase levels, the effect of phase quantization becomes negligible.
According to the inspection sheet supplied with the SLM, a $2\pi$ phase shift at $654$~nm corresponds to a grayscale value of approximately 130. This provides more than 32 addressable grayscale levels within the $0$--$2\pi$ phase range, satisfying the resolution requirement indicated by the numerical analysis. The resulting grayscale masks were displayed on the two corresponding regions of the SLM and experimentally evaluated using the setup shown in Fig.~\ref{fig:experimental-optical-setup}.

Figure~\ref{fig:optimized_residual_phase}(a) presents the optimization of the ODNN. 
The phase masks were optimized sequentially, with 1000 iterations performed for each mask. During optimization, the loss decreased rapidly in the early stage and gradually approached convergence, although noticeable fluctuations were observed between iterations.
Here, the signal is defined as the optical power within the designated detection regions, while the noise is defined as the remaining power recorded outside these regions.
The optimized coefficients are shown in Fig.~\ref{fig:optimized_residual_phase}(c). 
The optimized Zernike coefficients are distributed over multiple aberration modes.
In particular, the pronounced contribution of higher-order modes suggests that the discrepancy cannot be explained solely by simple alignment errors such as tip, tilt, or defocus, but also involves more complex wavefront distortions introduced by the folded optical path and the nonideal SLM implementation. Fig.~\ref{fig:optimized_residual_phase}(d) shows the the reconstructed residual phase profile for the first diffractive mask.
The optimized Zernike coefficients are distributed over multiple aberration modes excluding the piston term, the residual phase has a root-mean-square variation of $2.0135$~rad
at the operating wavelength. The correction contains contributions from different Zernike modes, indicating that the experimental distortion cannot be attributed to a single conventional aberration.  

\begin{figure}[htbp]
    \centering
    \includegraphics[width=0.7\textwidth]{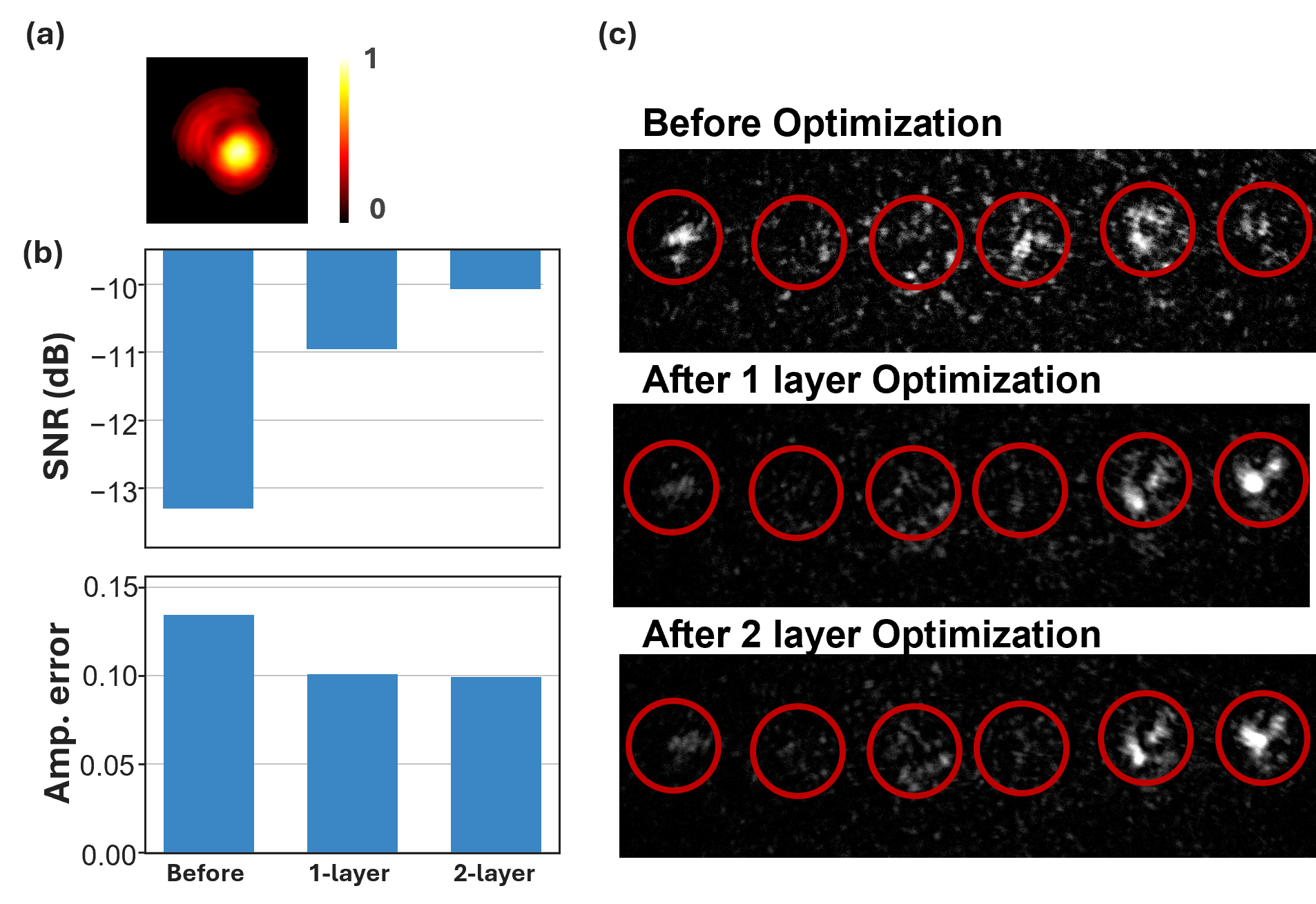}
    \caption{Experimental results before and after two-layer optimization.(a) System input, (b) SNR and amplitude error results, (c) Camera output. The SNR increases from $-13.3$~dB to $-10.04$~dB, corresponding to an improvement of approximately $3$~dB. The amplitude error reduces from 0.14 to 0.10. }
  \label{fig:results_exp}
\end{figure}

After optimization, the residual correction of each layer was superimposed on its numerically designed phase mask before being displayed on the SLM. Applying the optimized corrections to both diffractive layers increased the experimentally measured SNR from $-13.3$~dB to $-10.04$~dB, corresponding to an improvement of $3.26$~dB.
Fig.~\ref{fig:results_exp} summarizes the experimental optimization results. Fig.~\ref{fig:results_exp}(a) shows the actual input of the ODNN. In the experiment, the modal weights of the input field were measured by mode decomposition~\cite{zhang2026decomposing} and used as the ground truth. Fig.~\ref{fig:results_exp}(b) compares the performance before and after optimization. The most significant improvement is obtained after optimizing the first phase mask. This is mainly because the first layer acts directly on the incident field and simultaneously determines the optical field propagating toward the second layer. Therefore, its correction can compensate not only for aberrations present in the input field, but also for accumulated wavefront distortions affecting the field incident on the second phase mask.
Fig.~\ref{fig:results_exp}(c) provides a direct visualization of the ODNN outputs before and after optimization. A clear concentration of optical power within the target regions and a reduction of background leakage can be observed after optimization.
Notably, although the optimization loss was defined solely based on the SNR, the modal power distribution also became more accurate after optimization. As a results, the amplitude error decrease from 0.14 to 0.10.
This result demonstrates that low-dimensional Zernike optimization can effectively reduce the simulation-to-experiment discrepancy without re-optimizing the complete pixelated diffractive network.
The residual phase was inserted into the complete SLM phase composition
before display.

The SLM used in the experiment operates at an effective refresh rate of approximately 1 Hz, including the time required for a complete phase update and stabilization. The optimization speed is therefore mainly limited by the SLM update time and camera acquisition. In the present experiment, convergence was typically reached within 200 iterations, corresponding to a total optimization time of approximately 11 minutes. This optimization is required only once for system calibration.
Since the loss can fluctuate during the optimization, the phase profile corresponding to the lowest measured loss is retained as the final optimized solution.
The verified Zernike correction retained after SPGD is shown in
Fig.~\ref{fig:optimized_residual_phase}(d).

The experimentally measured output intensity distributions before and after hardware-in-the-loop optimization are shown in Fig.~\ref{fig:results_exp}(c). Before correction, considerable optical power was distributed outside the designated detection regions. After applying the optimized Zernike corrections to both layers, the target spots became more distinct and the background leakage was reduced. In this work, the measured total insertion loss is 8.14 dB, of which 3.22 dB is attributed to the finite reflection efficiency of the SLM. Since part of the incident light is directly reflected without being modulated by the SLM, a grating is added to the second phase mask to separate this background from the desired output. This filtering introduces additional loss. Therefore, the insertion loss could be further reduced by using an SLM with higher reflectivity at the operating wavelength and by improving the output-region design~\cite{bao2025real}.

\section*{Conclusion}

In conclusion, we demonstrated a programmable diffractive optical mode sorter implemented with an SLM-based architecture and introduced a low-dimensional hardware-in-the-loop optimization strategy to compensate for the discrepancy between numerical design and experimental implementation. Instead of re-optimizing the large number of pixel-level phase variables of the diffractive network, the residual system distortion was parameterized using a limited set of Zernike modes and optimized directly on the experimental setup using SPGD. This physics-informed correction substantially reduces the dimensionality of the experimental optimization while retaining the correct phase structure required for the target mode transformation. Using six LP modes of a MMF as a representative mode-sorting task, the proposed approach increased the experimentally measured SNR to an improvement of $3.26$~dB.
These results show that a significant fraction of the simulation-to-experiment gap in programmable diffractive optical systems can be compensated through low-dimensional adaptive calibration without retraining the complete diffractive network. 
It should be noted that folded optical architectures are commonly adopted in SLM-based multi-plane systems because they allow multiple phase-modulation stages to be realized with a single SLM. However, such folded configurations may introduce additional aberrations and alignment errors due to repeated reflections and off-axis propagation.

As a proof-of-concept demonstration, we experimentally implement a two-layer ODNN for sorting six spatial modes. The same framework can be scaled to a larger number of modes and improved performance by increasing the number of diffractive layers and trainable degrees of freedom~\cite{lib2025building}. 
More broadly, the proposed strategy provides a practical route toward self-calibrating and reconfigurable spatial-mode processors, with potential applications in adaptive mode multiplexing and demultiplexing, MMF communications~\cite{czarske2026complex}, programmable optical switching and routing, mode-resolved sensing, optical computing~\cite{cheng2026physical,cui2026universal}, biomedical imaging~\cite{kuschmierz2021ultrathin,sun2024lensless,sun2024projection}, secure communication~\cite{lin2025secret,li2026nonvolatile}, and high-dimensional classical and quantum optical information processing~\cite{zhang2026decomposing,lib2025high,krause2026quantum,amiri2026quantum}.


\section*{Acknowledgements}


This research was funded by the German Research Foundation for funding the Reinhart Koselleck project (project number: 560574412).

\section*{Conflict of interest}
The authors declare no competing interests.

\section*{Author contributions statement}
Qian Zhang and Shiyue Chen conceived the concept. 
Shiyue Chen and Qian Zhang conducted the experiments.
Shiyue Chen and Qian Zhang conducted the programming and the training.
Qian Zhang, Shiyue Chen, and Juergen Czarske analysed the results.
Juergen Czarske supervised the project.
All authors participated in discussions and contributed to the writing of the paper. All authors reviewed the manuscript. 

\section*{Data availability}
Data underlying the results presented in this paper are available from the corresponding author upon reasonable request.

\bibliography{sample}

\end{document}